\documentclass[doublecol]{epl2}

\usepackage{graphicx}
\usepackage{amsmath}
\usepackage{amsfonts}
\usepackage{amssymb}

\usepackage{color}
\usepackage{bm}      
\usepackage{array}

\usepackage{hyperref}  

\newcommand{\bea}{\begin{eqnarray}}
\newcommand{\eea}{\end{eqnarray}}
\newcommand{\beq}{\begin{equation}}
\newcommand{\eeq}{\end{equation}}
\newcommand{\bit}{\begin{itemize}}
\newcommand{\eit}{\end{itemize}}

\newcommand{\s}{\sigma}
\newcommand{\ot}{ \omega}
\newcommand{\D}{\mathrm{d}}
\newcommand{\+}{|+ \rangle}

\renewcommand{\arraystretch}{1.5}

\newcolumntype{C}{ >{\centering\arraybackslash} m{13.5cm} }

\begin{document}

\title{Modulated two-level system : Exact work statistics}

\author{Gatien Verley$^1$, Christian Van den Broeck$^2$, Massimiliano Esposito$^1$}

\institute{
  \inst{1} Complex Systems and Statistical Mechanics, University of Luxembourg, L-1511 Luxembourg, G.D. Luxembourg, EU \\
  \inst{2} Hasselt University - B-3590 Diepenbeek, Belgium, EU \\
}

\pacs{05.70.Ln}{Nonequilibrium and irreversible thermodynamics}
\pacs{05.40.-a}{Fluctuation phenomena, random processes, noise, and Brownian motion}
\pacs{05.20.-y}{Classical statistical mechanics}

\date{\today}

\abstract{We consider an open two-level system driven by a piecewise constant periodic field and described by a rate equation with Fermi, Bose and Arrhenious rates respectively. We derive an analytical expression for the generating function and large deviation function of the work performed by the field and show that a work fluctuation theorem holds.}

\maketitle

\section{Introduction}

According to standard thermodynamics, the amount of work $\langle W \rangle$ needed to bring a system in contact with an heat bath at temperature $T$ from one equilibrium state to another one, is at least the corresponding difference in equilibrium free energy $\Delta F^{\rm eq}$. This result is a direct consequence of the second law. Over the past two decades, this issue has been revisited for the case of driven non-macroscopic systems, with surprising theoretical consequences. For a small system, the work $W$ will fluctuate from one experiment to the other and the full distribution of work $P_W$, rather than solely the average, becomes the experimentally accessible quantity of interest. Starting from basic physical principles, one can derive the Jarzynski equality \cite{Jarzynski1997_vol78,Jarzynski2006_vol73} $\langle \exp(-\beta W)\rangle= \exp(- \beta\Delta F^{\rm eq})$ and the Crooks work theorem  \cite{Crooks2000_vol61,Chelli2009_vol130,Harris2007_vol2007,Chen2008_vol129,Lua2005_vol109,Speck2004_vol70} $P_W /\tilde{P}_{-W}=\exp\{\beta W- \beta \Delta F^{\rm eq} \}$ (the tilde referring to time-reversed driving) when considering a system initially at equilibrium with a single heat bath at inverse temperature $\beta = 1/(k_B T)$ (set to unity throughout the paper). These results imply the standard thermodynamic inequality $\langle W \rangle-\Delta F^{\rm eq} \geq 0$. In light of these tantalizing developments, there has been an obvious interest in verifying that distributions of work indeed satisfy these equalities, and if possible, to calculate their explicit form. This has been achieved in a number of scenarios including models with Langevin dynamics \cite{Fogedby2012_vol2012, Farago2002_vol107, Chakrabarti2009_vol72, Hoppenau2013_vol, Pal2013_vol87, Zon2003_vol67, Imparato2009_vol6, Trieu2007_vol75}, (granular) gas models \cite{Visco2006_vol125, Bena2005_vol71, Cleuren2006_vol96}, mean field models \cite{Imparato2005_vol70, Imparato2005_vol72} and a discrete (toy) model \cite{Kumar2011_vol84}. $P_W$ has also been measured experimentally \cite{Collin2005_vol437, Junier2009_vol102, Douarche2005_vol2005, Joubaud2009_vol2009, Andrieux2008_vol2008} and numerically \cite{Granger2010_vol2010,Einax2009_vol80,Saha2008_vol77}.

Somewhat surprisingly, the calculation of the work distribution is notoriously difficult for one of the prototype models of statistical mechanics, namely a two-state system driven by a modulated field \cite{Ritort2004_vol2004,Subrt2007_vol2007,Esposito2007_vol76}. Even the case of periodic modulation is challenging as it turns out to be, mathematically speaking, closely related to the parametric oscillator \cite{Adrianova1995_vol}. 
The main purpose of this letter is to provide an exact analytical solution for the work distribution of a two-level system subjected to periodic piecewise constant modulation. We show that they obey a work fluctuation theorem, discuss the similarities and differences between Fermi, Bose, and Arrhenius rates, and explore various limiting regimes. 

\section{Periodically Modulated system}
\label{sec:ModelDesc}

We consider a two-level system $\s = \pm 1$ subjected to an external field $h$ and coupled to a heat bath at temperature $T$. The energy change in the system obeys the first law of thermodynamics: the rate of change of the system energy $E= - h  \s$ is the sum of a work flow $\dot{W}=- \dot h \s$ and a heat flow $\dot{Q}=-  h \dot \s$, i.e. $\dot{E}=\dot{W}+ \dot{Q}$.
We will focus on the evaluation of the cumulated work
\beq
W = -\int_{0}^t \D t' \dot h (t') \s(t').  \label{eq:workDef}
\eeq

Due to its interaction with the heat bath, the system undergoes thermal transitions between its two states. Let $\omega_{-\s,\s}$ denote the probability per unit time for the system to flip from state $\s$ to state $-\s$. The resulting Markovian stochastic dynamics is characterized by a $2$ by $2$ transition rate matrix $\bm{L}$ with elements $L_{\s,\s'}= -\s \s' \omega_{-\s',\s'}$ where
\beq
\omega_{-\s,\s} = \omega(h) e^{-\s h}. 
\label{omega}
\eeq
These rates satisfy local detailed balance \cite{Seifert2012_vol75,Esposito2012_vol85}
\bea\label{eqd}
\frac{\omega_{-\s,\s}}{\omega_{\s,-\s}} =  \frac{p^{eq}_{-\s}}{p^{eq}_{\s}} \;\;\; \mbox{where} \;\;\; p^{eq}_\s=\frac{e^{\beta \s h}}{Z}.
\eea
The form (\ref{omega}) includes as special cases, Arrhenius rates $\omega(h) = \Gamma$, Fermi rates $\omega(h) = \Gamma/ (2 \cosh(h))$, and Bose rates $\omega(h) = \Gamma/ (2 |\sinh(h)|) $, where $\Gamma$ is a positive constant setting the time scale (and set to unity in our plots).

The rate of change of the work, $\dot W=- \dot h \s$, is a deterministic function of the process $\s$.  Hence the joint set of variables $(\s, W)$ again defines a Markov process, and the corresponding joint probability $P_{\s,W}$ obeys the following evolution equation
\beq
\partial_t P_{\s,W}  =  \sum_{\s'=\pm 1} L_{\s,\s'} P_{\s',W} -\partial_{W}\dot{W}P_{\s,W}. \label{eqev}
\eeq
with the initial condition, $P_{\s,W}(0) = \delta(W) p_\s(0) $, being the probability to be on state $\s$ with $W=0$ at time $t=0$.
The probability distribution for the cumulated work follows by summation over the system states
\bea
P_{W}=\sum_{\s=\pm 1} P_{\s,W}.
\eea
By introducing the generating function
\bea
G_{\mu}= \sum_\s G_{\s,\mu} \  \ {\rm where} \  \ G_{\s,\mu} =  \int_{-\infty}^{\infty} \D W e^{\mu W} P_{\s,W}, \label{eq:DefOfGenFunc}
\eea
one obtains from (\ref{eqev})
\beq
\partial_t G_{\s,\mu}= \sum_{\s'=\pm 1} L^{(\mu)}_{\s,\s'} G_{\s',\mu}, \label{eq:GenFuncEvo}
\eeq
with $\bm{L}^{(\mu)}$ a matrix with elements $L^{(\mu)}_{\s,\s'} = L_{\s,\s'}-\dot h \mu \s \delta_{\s,\s'} $. From now on, we assume that the perturbation $h=h(t)$, and hence also the matrix $\bm{L}^{(\mu)}$, is time-periodic. We can thus write the solution of (\ref{eq:GenFuncEvo}) after $n$ periods as 
\beq
 G_{\s,\mu}(n\tau)=\sum_{\s'=\pm 1} (Q^n)_{\s,\s'} p_{\s'}(0),
\eeq
The matrix $\bm{Q}$ is the single period propagator
\bea
\bm{Q} &=&  \overrightarrow{\exp} { \int_{0}^{\tau} \bm{L}^{(\mu)}(t) \D t}, \label{eq:factProp}
\eea
where $ \overrightarrow{\exp}$ stands for the time-ordered exponential.
Let $\lambda$ and $\lambda'$, where $\lambda \geq \lambda'$, denote the eigenvalues of $\bm{Q}$ with corresponding left and right eigenvectors $\langle \lambda|$,$\langle\lambda'|$ and $|\lambda\rangle$, $|\lambda'\rangle$, respectively. Since $\bm{Q}^n=\lambda^n|\lambda\rangle|\langle \lambda|+\lambda'^n|\lambda'\rangle|\langle \lambda'|$, the asymptotic behavior of $G_{\mu}(n\tau)$ for large times or large $n$ is determined by the largest eigenvalue $\lambda$ of $\bm{Q}$:
\bea
 \lim_{n \rightarrow \infty}\frac{1}{n}\ln G_{\mu}(n\tau)=\ln\lambda=\phi_\mu , \label{eq:DefOfGenFunc2}
 \eea
where
\bea
\lambda = \frac{  \mathrm{tr} \, \bm{Q} + \sqrt{ \left [ \mathrm{tr} \, \bm{Q}\right ]^{2} - 4 \det \bm{Q} }}{ 2} \label{eq:ScaledGenCumFunc}.
\eea
To investigate the asymptotic behaviour, we consider the work per period w = W/n  and focus on the large deviation function of $P_{W=n w}$ for $n \rightarrow \infty$ defined by
\beq
I_w=-\lim_{n \rightarrow \infty} \frac{1}{n} \ln P_{n w}.  \label{eq:defLDF}
\eeq
This function quantifies the probability of exponentially rare deviations of $W$ from its average value $\langle W \rangle =n \langle w \rangle $. 
\beq
P_{W=n w} \asymp e^{-n I_w}.
\eeq 
Note that $\langle w \rangle $ corresponds to the minimum of $I_w$ to ensure a long time convergence of the work $W$ toward $n w$ \cite{Touchette2009_vol478}.
Inserting expression (\ref{eq:defLDF}) in (\ref{eq:DefOfGenFunc}), one finds 
\bea
G_{\mu} \asymp \int_{-\infty}^{\infty} n \D w  \exp{ \left[ -n (I_w - \mu w ) \right]}.
\eea
Using the saddle point approximation and (\ref{eq:DefOfGenFunc2}) leads to  
\beq\label{lt}
\phi_\mu =  \max_{w}\left \{ \mu w - I_w \right \}.
\eeq
In words, $\phi_\mu$ and $I_w$ are related to each others by a Legendre transform. The evaluation of both the asymptotic generating function and the large deviation function are reduced to the calculation of the trace and determinant of the propagator $\bm{Q}$, cf. (\ref{eq:ScaledGenCumFunc}).

\section{Piecewise Constant Driving}

\begin{figure}[h]
\begin{center}
\includegraphics[width= 7cm]{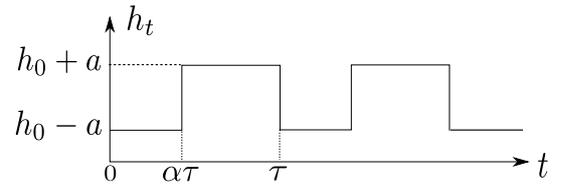} 
\end{center}
\caption{External field following a piecewise constant protocol oscillating between the values $h_0-a$ to $h_0 + a$ over a period of duration $\tau$. }
\label{fig0}
\end{figure}

In order to evaluating analytically the propagator $\bm{Q}$, we now consider the piecewise constant driving sketched on Fig.~\ref{fig0}. This driving is characterized by four parameters: the intensity of the field $h_0$, the amplitude of the jumps $ 2a$, the period of the driving $\tau$ and the cyclic ratio $\alpha$ (i.e. the fraction of time per period spent at the low value of the field). 

Despite its apparent simplicity, the piecewise constant driving displays discontinuities producing delta-function contributions in $\dot{h}$.  We explain in the appendix how to deal with the discontinuities of the protocol to obtain $\bm{Q}$ and thus the work statistics. The final result reads:
\bea
\mathrm{tr} \, \bm{Q} &=& A\cosh\{2a(2\mu+1)\}+B, \label{eq:traceG} \pagebreak \\
\det \bm{Q} &=&  C, \label{eq:detG}
\eea
where $A$, $B$ and $C$ are constants independent of $\mu$
\bea
A &=& \frac{ ( 1 - z^+)  (1 - z^-)}{\cosh 2a + \cosh 2h_0}, \label{eq:coeffA} \\
B &=& \frac{\left ( 1+z^+z^- \right ) \cosh 2h_0 +\left (z^++z^- \right)  \cosh 2a }{\cosh 2a + \cosh 2h_0}, \nonumber  \\
C &=& z^+z^-, \nonumber \\
\mbox{with} \pagebreak \nonumber \\
z^-  &=& \exp(-\alpha \tau \ot^-), \label{eq:zminus}\\
z^+ &=& \exp[-(1-\alpha) \tau \ot^+],\nonumber  \\
\ot^\epsilon &=& 2 \omega(h_0 + \epsilon a) \cosh(h_0 + \epsilon a)  \;\; \mbox{with} \;\; \epsilon =0,\pm \nonumber .
\eea
The crucial point to note is that the dependence on $\mu$ via the expression $\cosh\{2a(2\mu+1)\}$ is relatively simple. This feature can be exploited when performing the Legendre transform $ I_w = \max_\mu \left \{ \mu w -\phi_\mu \right \} $ and leads to (see appendix for details) 
\begin{equation}
I_w = -\frac{w}{2} + \ln \left [ \frac{2\left (x_w+\sqrt{(x_w)^2-1}\right )^{|w|/4a}}{Ax_w+B+ \sqrt{(Ax_w+B)^2-4C}} \right ], \label{eq:firstMainResult}
\end{equation}
with 
\begin{multline}
x_w = -\frac{Bw^2}{A(w^2-16a^2)} \label{eq:xFunctionOfw} \\ 
-\frac{\sqrt{B^2w^4 -  (w^2-16a^2)\left [w^2(B^2-4C)+16A^2a^2\right ]}}{A(w^2-16a^2)}.
\end{multline}
The explicit expressions for the asymptotic work generating function, (\ref{eq:ScaledGenCumFunc}) with (\ref{eq:traceG}) and (\ref{eq:detG}), and for the large deviation function, (\ref{eq:firstMainResult}), are the main results of this paper. 

\section{Discussion}

\begin{figure}
\begin{center}
\includegraphics[width= 8.5cm]{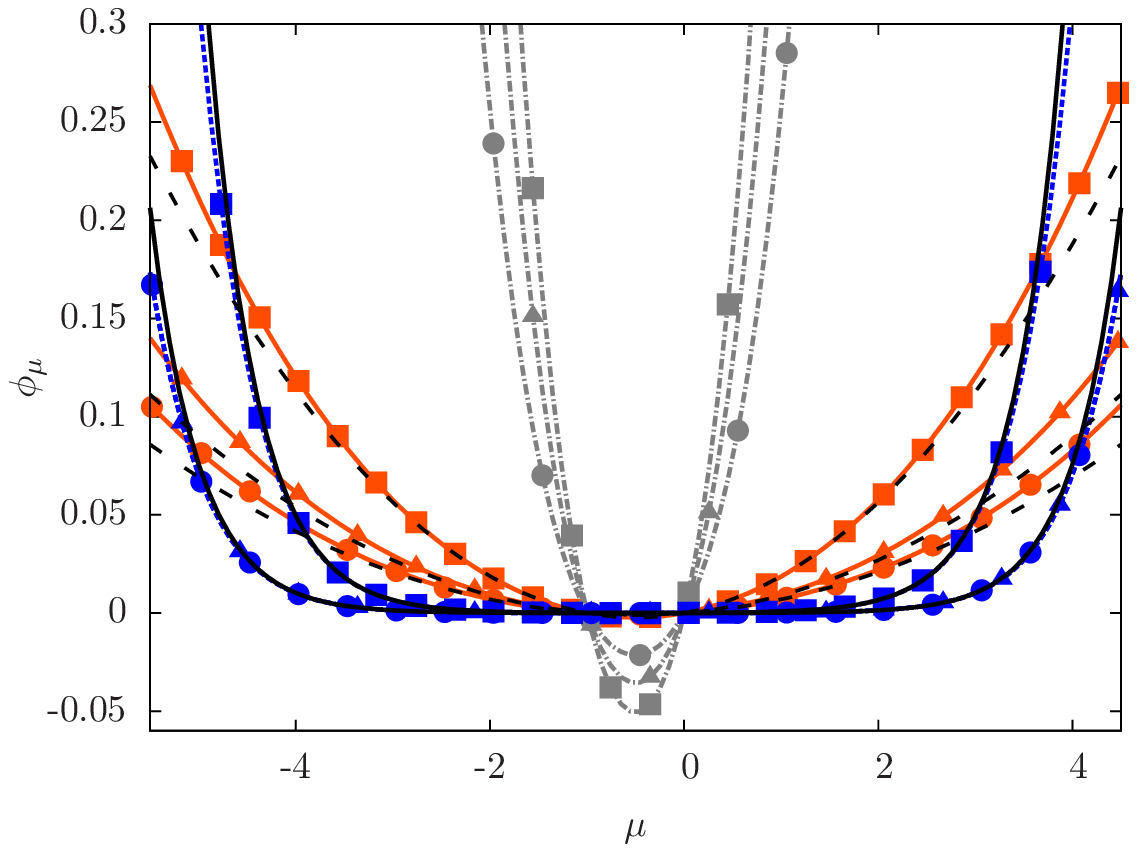}
\includegraphics[width= 8.5cm]{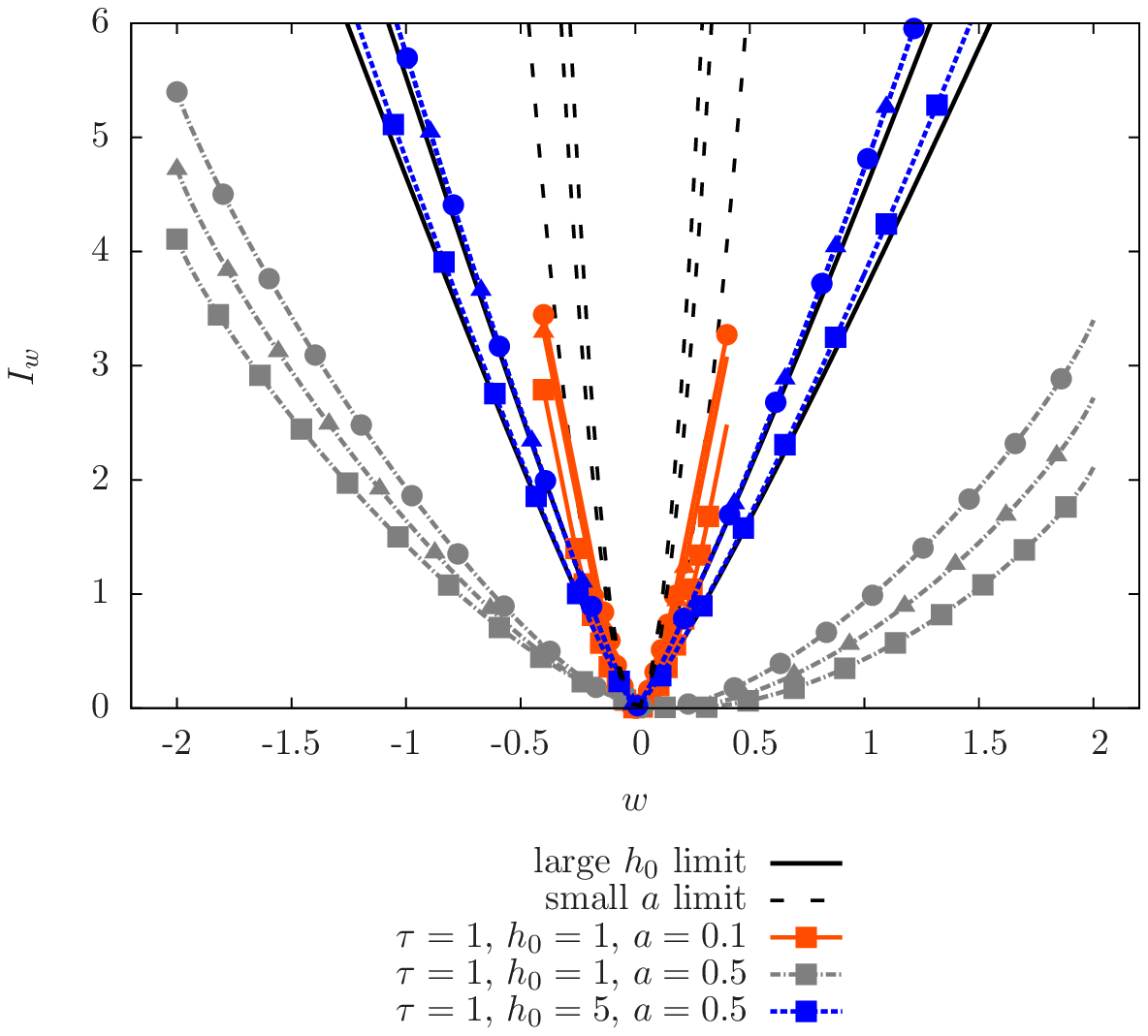}
\end{center}
\caption{(Top) Cumulant generating function $\phi_\mu$ of work per period as a function of the Laplace parameter $\mu$. (Bottom) Large deviation function $I_w$ versus work per period. Various values of the field are plotted: high field $h_0=5$ and $a=0.5$ (blue dashed line), intermediate field $h_0 = 1$ and $a=0.5$ (grey dotted dashed line), and low amplitude of the driving $h_0=1$ and $a=0.1$ (orange solid line). Symbols encode the types of rates: Arrhenius (squares), Bose (triangles) and Fermi (circle). The other parameters are $\tau =1$, and $\alpha=0.3$.}
\label{fig3and4}
\end{figure}

\begin{figure*}
\begin{center}
\includegraphics[width= 8.5cm]{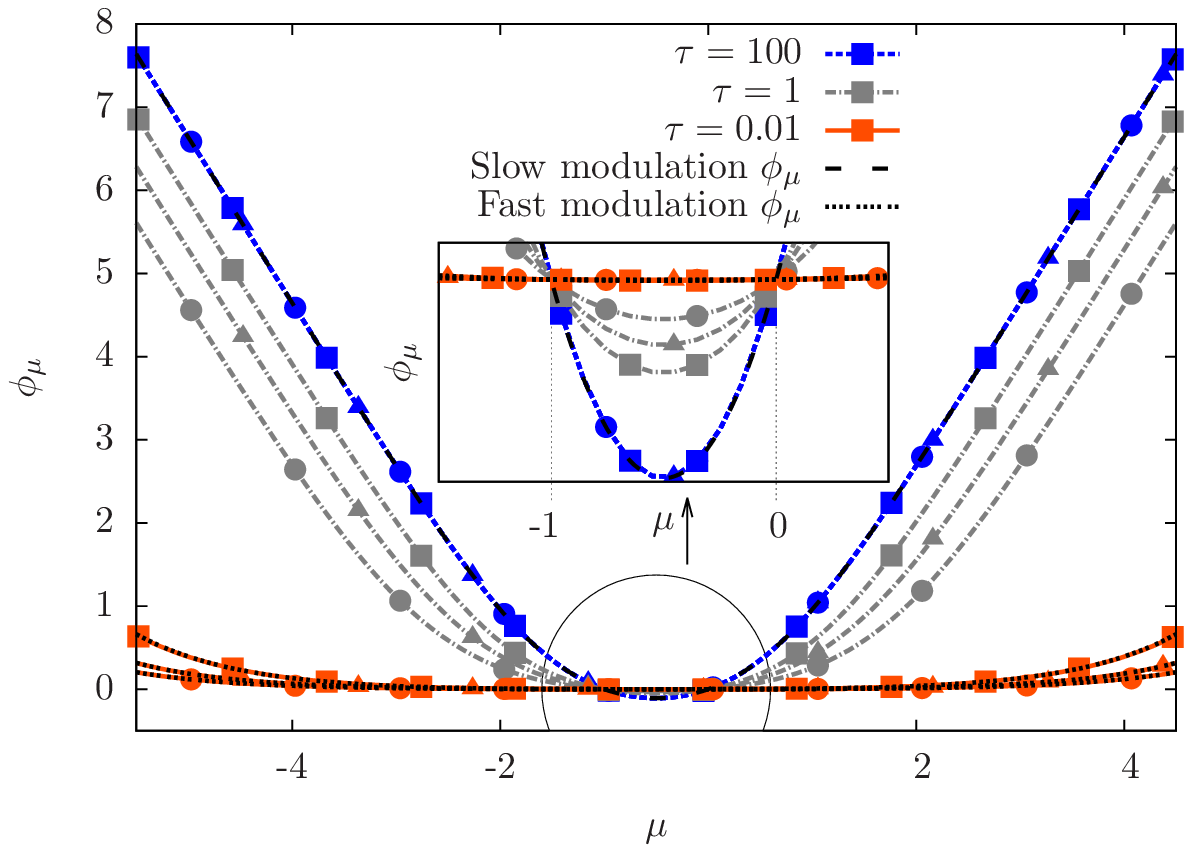}
\includegraphics[width= 8.5cm]{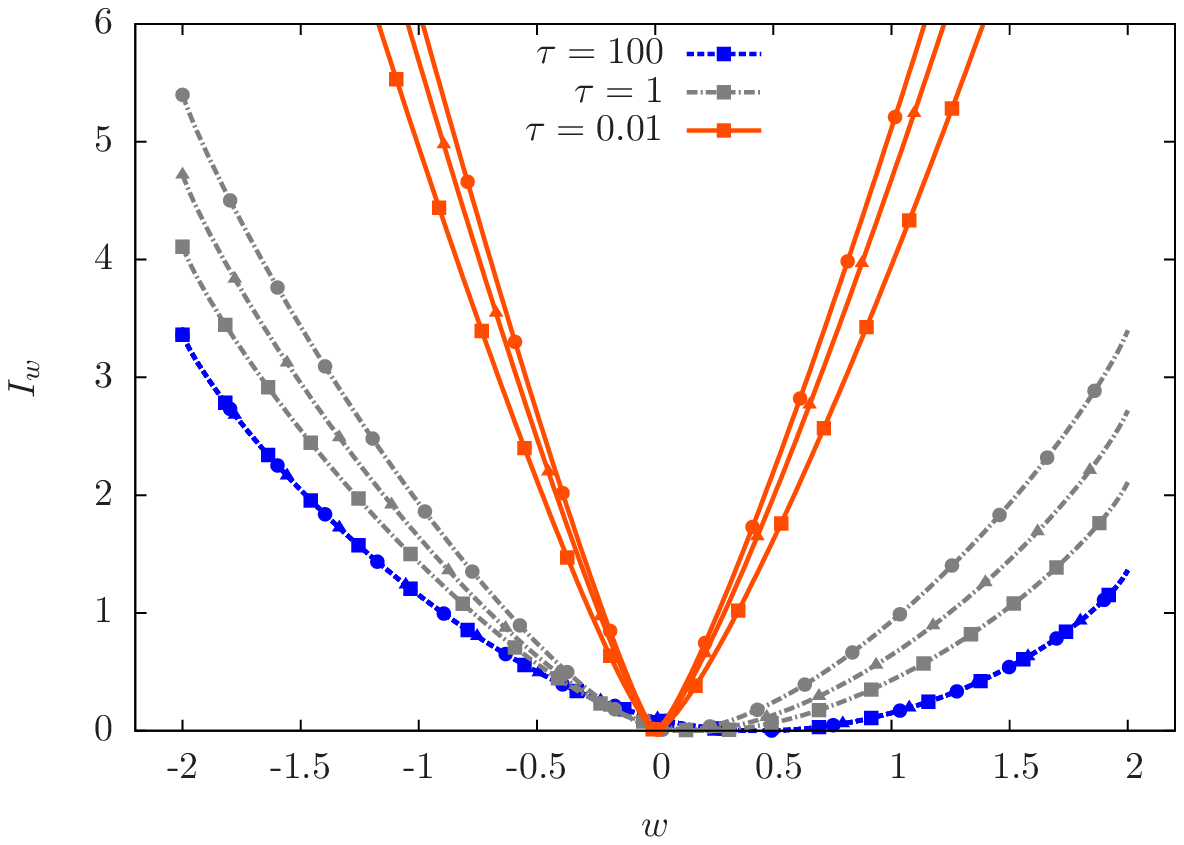}
\end{center}
\caption{(Left) Cumulant generating function $\phi_\mu$ for the work per period versus the Laplace parameter $\mu$. (Right) Large deviation function $I_w$ versus work per period for various characteristic time scales of the driving: $\tau=100$ (blue dashed line), $\tau=1$ (gray dotted dashed line) and $\tau = 0.01$ (orange solid line). Same symbols code as in Fig.~\ref{fig3and4}. Here $a=0.5$, $h_0=1$ and $\alpha=0.3$.}
\label{fig1and2}
\end{figure*}

In its traditional formulation, the Crooks fluctuation theorem applies to systems initially at equilibrium and is valid for any time. It connects the work fluctuations arising when applying an arbitrary protocol to those of a different experiment where the time-reversed protocol is considered. This result is a special case of the universal detailed fluctuation theorem for entropy production \cite{Seifert2005_vol95,Esposito2010_vol104,Verley2012_vol108,Verley2012_vol86a,Broeck2013_vol}. Indeed, when a system is in contact with a single reservoir, entropy production is given by the work minus the change in nonequilibrium free energy of the system. This latter reduces to the difference of  equilibrium free energy in the traditional Crooks formulation. For a periodic driving, the change in nonequilibrium free energy over a period becomes zero when initial transients are gone and thus plays no role in the long time limit. 
Furthermore, for our piecewise constant modulation, the time-reversed driving is identical to the forward driving up to a time-shift which again plays no role in the long time limit. As a result, the detailed fluctuation theorem for entropy production becomes a Crooks-like work fluctuation theorem for long times of the form ${P_W }/{{P}_{-W}}=\exp\{ W\}$. More precisely, the large deviation function and the work generating function satisfy the fluctuation theorem symmetry
\bea
I_w-I_{-w}=-w, \label{FT_LD}\\ 
\phi_\mu=\phi_{-1-\mu}. \label{FT_GF}
\eea
These relations are easily verified. The second term in the large deviation function (\ref{eq:firstMainResult}) is even in $w$, and the first term immediately reproduces (\ref{FT_LD}).
For the work generating function, the $\mu$ dependency of $\phi$ appears only through the function $\cosh[2a(1+2\mu)]$ in (\ref{eq:traceG}) which is indeed invariant under the exchange of $\mu$ with $-1-\mu$. The detailed fluctuation theorem implies a Jarzynski-like integral fluctuation theorem which for the generating function reads $\phi_{-1}=1$. Both the detailed and the integral fluctuation theorem are satisfied on Figs.~\ref{fig3and4}-\ref{fig1and2}, where $\phi_\mu$ and $I_w$ are plotted for various values of the protocol parameters, and for Arrhenius, Fermi and Bose rates. 

\begin{figure}
\begin{center}
\includegraphics[width= 8.5cm]{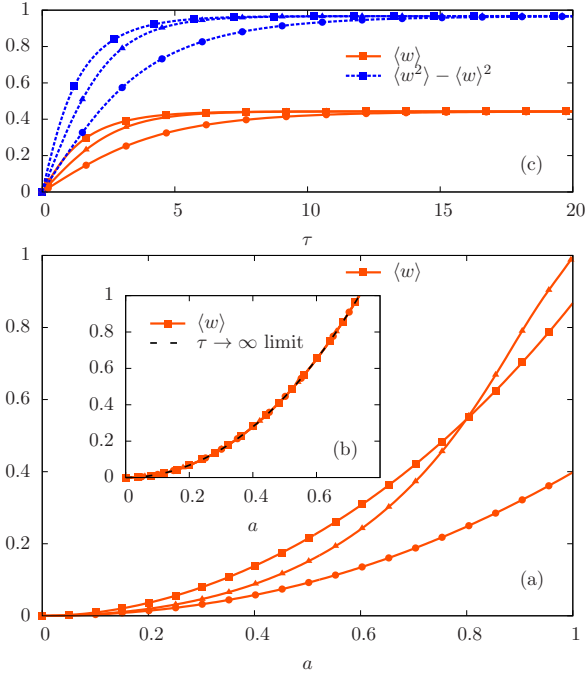}
\end{center}
\caption{Average and variance of work per period as a function of the modulation amplitude $a$ for (a) and (b) or as a function of the period $\tau$ for (c). Parameters are  $\alpha =0.3$, $h_0=1$, with $\tau = 1$ for (a), $\tau =100$ for inset (b) and $a=0.5$ for (c). The symbol code is the same as in Fig.~\ref{fig3and4}.} 
\label{fig5}
\end{figure}

These plots reveal a number of other features, which can be verified via analytical calculations (see Table~\ref{tab:limitcases}). 
First, due to the finite support of the large deviation function, the generating function displays a linear asymptotic behaviour for $\mu \rightarrow \pm \infty$. The physical origin of this finite support is the existence of an upper and lower bound for the work per period, namely $\pm 4a$ ($\pm 2a$ for every jump in the field). 
Second, the work variance (i.e. the width of the large deviation function) typically increases as the average number of jumps per period increases, see Fig.~\ref{fig5}: we observe, as the average number of jumps decreases from Arrhenius over Bose to Fermi rates, a corresponding decrease in the variance.
Third, in the limit of infinite period, $\tau \rightarrow \infty$, the system has time to relax to the prevailing equilibrium distribution after each jump in the field. In this case, the work distribution becomes independent of the types of rates $\omega(h)$ since they all lead to the same equilibrium distribution, cf. (\ref{eqd}). 
Since the field undergoes jumps, we are however not in a close-to-equilibrium regime. 
Fourth, the latter regime is reached in the limit of small jumps $a \rightarrow 0$ where the work distribution becomes Gaussian
\beq
I_w = \frac{(w-\langle w \rangle )^2}{4 \langle w \rangle }. \label{eq:GaussianWorkFluct}
\eeq
The fact that the variance equals twice the average work is the signature of the fluctuation theorem for Gaussian processes. The general form of the average work is given by 
\beq
\langle{w}\rangle = \frac{4a \sinh(2a)\left( 1-e^{-(1-\alpha)\tau\ot^+} \right) \left( 1-e^{-\alpha \tau\ot^-} \right)}{\left( \cosh(2h_0) + \cosh(2a) \right)\left( 1-e^{-\alpha \tau\ot^- -(1-\alpha)\tau\ot^+} \right)}, \label{eq:AvgWorkPer}
\eeq 
and is plotted on Fig.~\ref{fig5}. In the close to equilibrium regime, it becomes a quadratic function in the perturbation amplitude $a$.
Fifth, a far from equilibrium regime is reached in the limit of fast modulation $ \tau \rightarrow 0$ or large field $h_0 \rightarrow \infty$. In both cases, the number of spin flips per period become small and the work distribution converges to a delta function corresponding to a vanishing work per period.

\section{Conclusion}

The two-level system has played a crucial role in statistical physics to reveal the properties of both equilibrium and non-equilibrium systems. In the present letter, we derived an exact analytic expression for the (asymptotic) work distribution and work generating function of a two-level system in contact with a single thermal heat bath and subjected to a periodic piecewise constant field. We also showed that the universal fluctuation theorem for entropy production reduces to a corresponding Crooks-like fluctuation theorem for work. Our study could be easily extended to more complicated situations. The case of several heat bath is of obvious relevance since it would allow to discuss the way in which the field modifies the energy transfers between the various baths. One could also increase the number of field states in the piecewise driving to break the asymptotic time-reversal symmetry $P=\tilde{P}$ of the present study.  

\renewcommand{\arraystretch}{2.4}
\setlength{\tabcolsep}{0.4cm}
	\begin{largetable}
		\hspace{-1cm} \begin{tabular}{c|C} 
$\Gamma \tau \ll 1$ & \raisebox{0.12cm}{$ 
\frac{ \tau}{2} \left [ -\alpha \ot^- -(1-\alpha)\ot^+ +\sqrt{ \left ( \alpha \ot^- +(1-\alpha)\ot^+ \right )^2 + 4 \alpha (1-\alpha) \ot^+ \ot^- \frac{\cosh 2a(2\mu+1) - \cosh 2 a }{ [ \cosh 2a + \cosh 2 h_0 ]}  } \right]$ } \\
$\Gamma  \tau \gg 1$ & \raisebox{0.12cm}{$ \ln \frac{ \cosh 2 h_0 + \cosh 2a(2\mu+1)}{\cosh 2a + \cosh 2 h_0} + \left (e^{-(1-\alpha) \tau \ot^+} + e^{-(1-\alpha) \tau \ot^+} \right ) \frac{ \cosh 2 a - \cosh 2a(2\mu+1)}{\cosh 2h_0 + \cosh 2a(2\mu+1) }$}  \\
$ a \ll 1$ & \raisebox{0.12cm}{$ \mu (1+\mu) \langle w \rangle  =  8a^2 \mu (1+\mu) \frac{(1- e^{-(1-\alpha)\tau\ot^0})(1-e^{-\alpha \tau \ot^0})}{(1-e^{-\tau \ot^0})(1+\cosh 2 h_0)} $ }  \\
\raisebox{-0.6cm}{$h_0 \gg a$} & 
			\begin{tabular}{cc}
Arrhenius rates & Fermi and Bose rates \\
$\displaystyle \frac{\cosh 2a(2\mu+1) - \cosh 2a}{\cosh 2h_0} $  & $ \frac{\left (1-e^{-(1-\alpha)\Gamma \tau}\right ) \left (1-e^{-\alpha \Gamma \tau}\right )}{1-e^{-\Gamma \tau}} \frac{\cosh 2a(2\mu+1) - \cosh 2a}{\cosh 2h_0}  $ 
			\end{tabular} 
		\end{tabular}
\caption{Cumulant generating function of work per period $ \phi_\mu $ in various limits: Fast and slow modulation of the driving field ($\Gamma \tau \ll 1$ and $\Gamma \tau \gg 1$), small amplitude of change in the field ($a \ll 1$ ), and large values of the field ($h_0 \gg 1 $). Most of these expansions are valid when the Laplace parameter $\mu$ remains inside a given interval depending on the expansion parameter.  \label{tab:limitcases}}
	\end{largetable}

\renewcommand{\arraystretch}{1.4}
\setlength{\tabcolsep}{0.4cm}

\section{Appendix}

We denote the propagator for the system probabilities $p_\s(t)$ by
\beq
\bm{U}(t,t_0) =  \overrightarrow{\exp} { \int_{t_0}^{t} \bm{L}(t') \D t'}. \label{eq:freeProp}
\eeq
Then, the generating function over a period reads
\beq
\bm{Q} = \bm{K}^{(-2a)} \bm{U}(\tau ,\alpha \tau) \bm{K}^{(2a)} \bm{U}(\alpha \tau,0), \label{eq:propOverAPeriod}
\eeq
where $\bm{K}^{( \pm 2a)}$ is the operator cumulating the work over the jumps of amplitude $\pm 2a$ in the protocol. $\bm{K}^{( \pm 2a)}$ is obtained by taking the time ordered exponential of the generator $\bm{L}^{(\mu)}$ between time $t^-$ and $t^+$ just before and after a jump at time $t$,
\beq
\bm{K}^{(\pm 2a)} =  \overrightarrow{\exp} { \int_{t^-}^{t^+} \left (\bm{L}(t')-\mu \bm{\s_z}\dot h(t') \right ) \D t'},
\eeq
with $(\s_z)_{\s,\s'} = \s \delta_{\s,\s'}$ and $\dot h(t') = \pm 2a \delta(t-t')$. For $t^-$ infinitely close to $t^+$, the propagator over a jump of amplitude $h$ is 
\beq
K^{(h)}_{\s,\s'} = e^{-\s \mu h} \delta_{\s,\s'}. \label{eq:defOpX}
\eeq
At a constant value  of the field $h=h_0+\epsilon a$ with $\epsilon = \pm 1$, the propagator in equation (\ref{eq:propOverAPeriod}) simplifies to
\begin{multline}
U_{\s,\s'}(t,t_0) = \frac{e^{\s(h_0+\epsilon a)}}{2\cosh (h_0+\epsilon a)} \left ( 1 - e^{-\ot^\epsilon (t-t_0)}\right ) \\+ \delta_{\s,\s'} e^{-\ot^\epsilon(t-t_0)} . \label{eq:Usimplied}
\end{multline}
Using (\ref{eq:propOverAPeriod}), (\ref{eq:defOpX}) and (\ref{eq:Usimplied}), 
a lengthy calculation leads to $\bm{Q}$ and to the final explicit expression of the trace given in (\ref{eq:traceG}). 

The determinant of the propagator over a period is much easier to obtain \cite{Adrianova1995_vol}. If we define 
\beq
\bm{Q}(t) =  \overrightarrow{\exp} { \int_{0}^{t} \bm{L}^{(\mu)}(t') \D t'},
\eeq
we see that $\det \bm{Q}(t)$ obeys a closed evolution equation
\beq
\partial_t \det \bm{Q}(t) = \mathrm{tr} \,(\bm{L}^{(\mu)}(t)) \det \bm{Q}(t),
\eeq
which leads to 
\beq
\det \bm{Q}  = \exp \left ( \int_0^{\tau} \mathrm{tr} \, \bm{L}^{(\mu)}(t) \D t \right ) = z^-z^+.
\eeq

We now turn to the large deviation function, i.e. to the Legendre transform of the cumulant generating function
\beq
\phi_x = \ln \left[ Ax+B+\sqrt{(Ax+B)^2-4C}\right ] - \ln 2,
\eeq
where the $\mu$ dependence is hidden in the variable $x = \cosh 2a(2\mu+1)$. Hence, the Legendre transform $I_w =  \max_{\mu}\left \{ \mu w - \phi_\mu \right \}$ can be replaced by the extremum calculation in terms of $x$,   $I_w =  \max_{x}\left \{ \mu_x w - \phi_x \right \}$ where 
$\mu_x = \pm {\mathrm{argcosh} \, x}/{4a}-{1}/{2}.$
The resulting equation for $x $ as a function of $w$  turns out to be quadratic, with the proper solution for x given in (\ref{eq:xFunctionOfw}). The large deviation function (\ref{eq:firstMainResult}) is found by evaluating $\mu_x w -\phi_x$ in $x_w$ using the logarithmic representation of the hyperbolic cosine function.

\bibliographystyle{eplbib}
\bibliography{Ma_base_de_papier_no_url.bib}

\end{document}